\documentclass{emulateapj}
\usepackage{graphicx,amsmath,natbib}
\usepackage{epsf}
\usepackage{epstopdf}

\input{epsf}
\usepackage{epsfig}
\usepackage[countmax]{subfloat}
\usepackage{amsfonts}
\DeclareGraphicsExtensions{.jpg,.pdf,.png,.eps,.ps}

\def\um{{\hbox {$\mu$m}}}

\def\msol   {\ifmmode{{\rm M}_{\odot} }\else{M$_{\odot}$}\fi}
\def\lsol   {\ifmmode{{\rm L}_{\odot}}\else{${\rm L}_{\odot}$}\fi}

\slugcomment{}
\shorttitle{Size Bias and Differential Lensing of DSFGs}
\shortauthors{Hezaveh, Marrone \& Holder}
\begin{document}

\title{Size bias and differential lensing of strongly lensed, dusty galaxies identified in wide-field surveys}
\author{Yashar D. Hezaveh\altaffilmark{1}, Daniel P. Marrone\altaffilmark{2}, Gilbert P. Holder\altaffilmark{1}}
\altaffiltext{1}{Department of Physics,
McGill University, 3600 Rue University, 
Montreal, Quebec H3A 2T8, Canada}
\altaffiltext{2}{Steward Observatory, University of Arizona, 933 North Cherry Avenue, Tucson, AZ 85721, USA}

\begin{abstract}
We address two selection effects that operate on samples of
gravitationally lensed dusty galaxies identified in millimeter- and submillimeter-wavelength surveys.
First, we point out
the existence of a ``size bias'' in such samples: due to finite source
effects, sources with higher observed fluxes are increasingly biased
towards more compact objects. Second, we examine the effect of
differential lensing in individual lens systems by modeling each
source as a compact core embedded in an extended diffuse halo.
Considering the ratio of magnifications in these two components, we find that at
high overall magnifications the compact component is amplified by a
much larger factor than the diffuse component, but at intermediate
magnifications ($\sim$10) the probability of a larger magnification
for the extended region is higher. Lens models determined from multi-frequency resolved imaging data are crucial to correct for this effect.
\end{abstract}

\keywords{gravitational lensing ---
galaxies: luminosity function, mass function---
galaxies: abundances---
methods: numerical
}

\section{Introduction}

Surveys of the extragalactic sky at millimeter and submillimeter wavelengths have 
identified an optically dim galaxy population
in which rapid, obscured star formation powers far infrared (FIR) luminosities occasionally exceeding $10^{13}$~\lsol\ 
\citep[e.g.,][]{smail:97,barger:98,hughes:98,eales:99}.
 Surveys of these sources at 850 um using SCUBA were confusion-limited, with several thousand sources per square degree at mJy-level fluxes \citep{coppin06}. The source density falls steeply with increasing flux: at 850 ~\um\ there are only a handful of sources per square degree above 10 mJy \citep{coppin06}. In the first millimeter-wavelength survey to cover tens of square degrees with similar sensitivity, \citet{vieira:10} found that the rarest objects ($<1\,\mathrm{deg}^{-2}$) are far more abundant than would be expected from the extrapolation of lower flux density sources. 
Such a population is a natural consequence of the
gravitational lensing of intrinsically fainter galaxies
by intervening galaxies and galaxy clusters \citep{negrello:07,hezaveh:11, bethermin:11}, and thus the selection of very bright submillimeter sources from large area surveys has proven to be an efficient way to identify highly-magnified 
objects \citep{negrello:10}.

The gravitational magnification of these sources allows them to be examined 
in more detail:
they are seen at improved source-plane resolution (due to the magnification
from gravitational lensing), 
allowing the study of the star formation process on 
the scale of giant molecular clouds \citep{Swinbank:10}; furthermore, the
overall increase in flux allows higher signal-to-noise measurements
of the spectral energy distribution (SED) and molecular lines \citep{riechers:10, lestrade:11,frayer:11, Cox:11, Lupu:10}, 
which can be used as diagnostics of the gas conditions. 

Because gravitational lensing is a geometric effect, a given
position in the lensed object will be magnified equally at all wavelengths. 
However, the finite 
extent of the background galaxy will result in variations in the magnification applied to different regions within the galaxy
\citep{blandford:92};
the observed SED (e.g., \citealt{Blain:99}), as well as ratios of spectral lines 
\citep[e.g.,][]{downes:95}, will be distorted by this differential 
magnification if there are spatial variations in the physical conditions
within the source. 

The interpretation of observations of gravitationally magnified submillimeter galaxies (SMGs) are 
necessarily complicated by this possibility of differential magnification.
\citet{Blain:99} 
explained the excess mid-IR emission in some lensed SMGs
through preferential magnification of compact hot regions, 
which can flatten the observed spectrum at wavelengths shorter than the SED 
peak and thereby increase the inferred temperature and luminosity of the 
source. 
As shown below, a variety of effects are possible, with the possibility of
preferentially magnifying either compact or diffuse regions, depending
on the relative position of source and lens. 

The sensitivity advantages associated with observing lensed sources will make these galaxies prime targets for studying the physics of SMGs. 
However, biases introduced by 
the selection of lensed galaxies must be considered carefully when extrapolating physical properties (e.g., temperatures, luminosities, 
sizes) determined from these samples to the SMG population as a whole.

In this paper we consider two effects that operate in lensed SMG populations selected by their high flux in millimeter and submillimeter surveys. First, we use population models for SMGs, previously published in \citet{hezaveh:11}, to examine the total magnification present in such samples and point out the possibility of a strong size bias that such selection introduces. Second, we examine the differential 
magnification of a two-component galaxy model as a function of 
properties of the source, lens, and the source-lens alignment. 

For all calculations below, we assume a spatially flat universe with 
$\Omega_m=0.222$, $h=0.71$, $n_s=0.96$, 
and $\sigma_8=0.801$.

\section{Model parameters}
\label{sec:models}
\subsection{Population Modeling}
\label{sec:popmodels}
\citet{hezaveh:11} reproduced the observed abundance of bright SMGs at 1.4~mm by combining models of the redshift distribution of SMGs \citep{lacey:10,marsden:11} and a population of elliptical lenses with masses described by the \citet{sheth:99} mass function. For our examination of the population of lensed SMGs, we 
adopt the \citet{hezaveh:11} model, with the underlying SMG population described by the semi-analytic model of \citep{lacey:10}. The lensing magnification probability distribution is calculated as a function of source size by numerically calculating the lensing cross-section for a particular source size. \citet{hezaveh:11} assumed a mean source size and hence used a single lensing probability which was applied to the number counts. In this work we are examining the effect of the natural size distribution of the SMGs on the observed number counts and hence apply four lensing probability functions corresponding to four different source sizes. We simply assume a source population consisting equally of point sources (small enough to avoid magnification damping effects at the 1-100 range for galaxy mass lenses), and sources with radii of 1.0, 3.0, and 8.0 kpc and study their apparent relative abundance after lensing. Such sizes have been measured in resolved SMGs \citep[e.g.,][]{tacconi:08}.

\begin{figure}[h]
\centering
\makebox[0cm]{\includegraphics[trim=1mm 1mm 1mm 1mm, clip, scale=0.55]{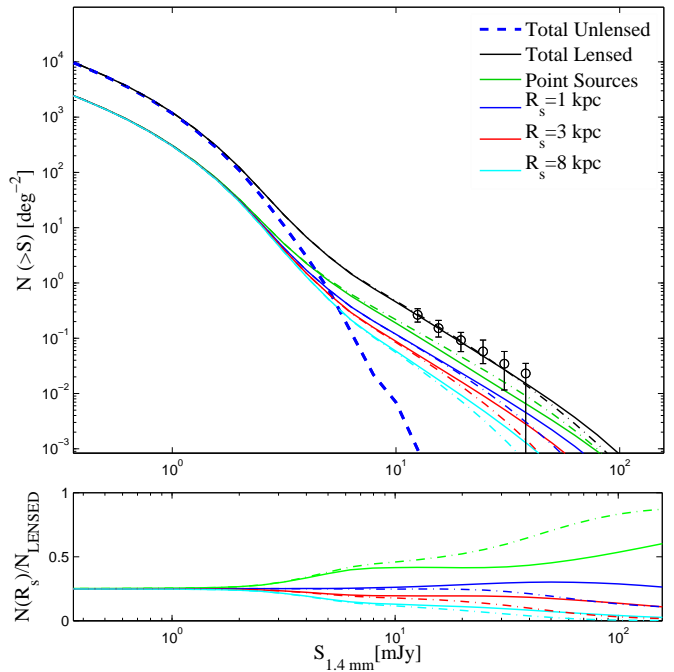}}
\caption{Lensed number counts of 1.4 mm (220 GHz) sources based on Durham semi-analytic model (\citealt{lacey:10}, as described in \citealt{hezaveh:11})
assuming four different source radii. 
The solid lines are calculated for a lens population with an ellipticity (1-axis ratio) of $\epsilon=0.1$, while the dot-dashed lines are for $\epsilon=0.3$. 
The lensing mass distribution is renormalized for the two cases so that the SPT (IRAS-excluded) dusty galaxy source counts 
(black circles; \citealt{vieira:10}) are matched.
The bottom panel shows the ratio of the lensed counts of each population to the total lensed number counts 
showing that at higher observed fluxes there is a strong bias towards selecting the most compact sources
from the input population.}
\label{f:sizes}
\end{figure}

\begin{figure}[h]
\centering
\makebox[0cm]{\includegraphics[trim=1mm 1mm 1mm 1mm, clip, width=8.5cm]{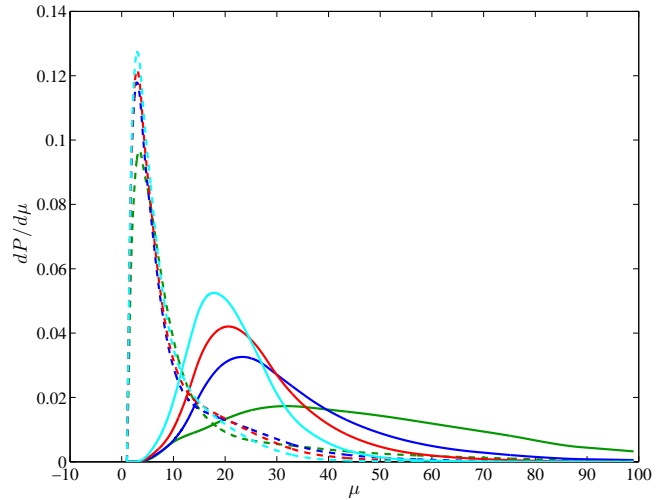}}
\caption{From the source count model of Figure~\ref{f:sizes} the distribution of source magnifications in the population of SMGs at two 
different observed flux densities. Green, blue, red, and cyan curves (bottom to top) correspond to point sources, and sources with radii of 1, 3, and 8 kpc, respectively. Solid curves show the magnification distributions at observed flux cut of 25 mJy and the dashed curves show the distribution at 2.5 mJy.  
For typical SMG spectra, these observed flux densities roughly correspond to 
350 
 and 35~mJy for {\it Herschel} surveys selecting sources at $\lambda$=500~\um. Unmagnified sources are excluded. 
The high magnification sources are an increasing fraction of the sources at the highest flux densities.}
\label{f:dpdmu}
\end{figure}

\subsection{Lens Models}
\label{sec:smgmodel}
For the purpose of studying the effects of differential lensing of a single lens system we use our ray-tracing code to simulate lensed images of extended sources. 
We model our lens as a Singular Isothermal Ellipsoid defined by its mass, ellipticity and angle and place the lens at $z_d=0.5$. 
The source is modeled with a slightly elliptical morphology as a two component model consisting of an extended component and an inner core with the ratio of the outer radii defined as $\alpha_R=R_{core}/R_{ext}$. 
The unlensed flux ratio of the two components is defined as $\beta_\mathrm{F}=F_{core}/F_{ext}=1$ and the total flux $F_T$ 
is simply the sum of the two component fluxes. After placing this source behind a lens the total magnification $\mu_\mathrm{T}$ is defined as the ratio of the lensed to unlensed flux of the source and $\beta_\mathrm{F}'$ is the observed flux ratio of the two components.

\begin{equation}
\mu_\mathrm{T}=\frac{\mu_{core} \, F_{core} + \mu_{ext} \, F_{ext}}{F_{core}+F_{ext}}=\frac{\beta_F \mu_{core}+\mu_{ext}}{\beta_F+1}
\label{eq1}
\end{equation}

\begin{equation}
\beta_\mathrm{F}'=\frac{\mu_{core}\, \,F_{core}}{\mu_{ext} \, \, F_{ext}}=\frac{\mu_{core}}{\mu_{ext}}\, \beta_F
\end{equation}

This morphology is motivated by the star forming structures observed in galaxies, which typically include large reservoirs of cold dust (extended component) in addition to small clumps of hot dust (core) in the vicinity of star forming regions \citep{pohlen:10,haan:11}. Although real galaxies may host multiple clumps with spatial offsets relative to the cold dust, this simple model can demonstrate the systematic effects caused by differential lensing of more complex sources.

\begin{figure*}[h]
\centering
\begin{minipage}[t]{0.48\linewidth}
\centering
\makebox[0cm]{\includegraphics[width=8.5cm]{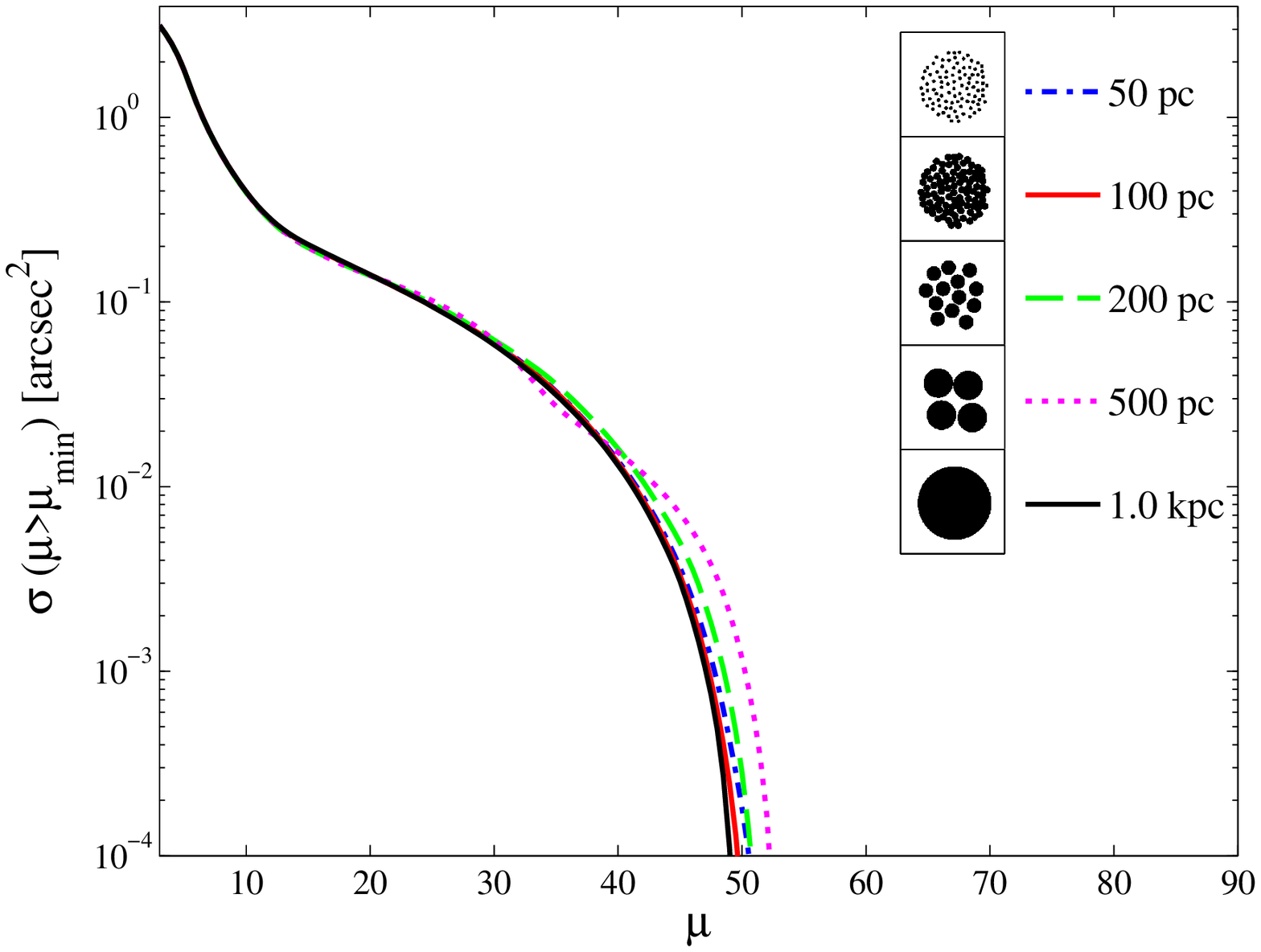}}
\end{minipage}
\begin{minipage}[t]{0.48\linewidth}
\centering
\makebox[0cm]{\includegraphics[width=8.5cm]{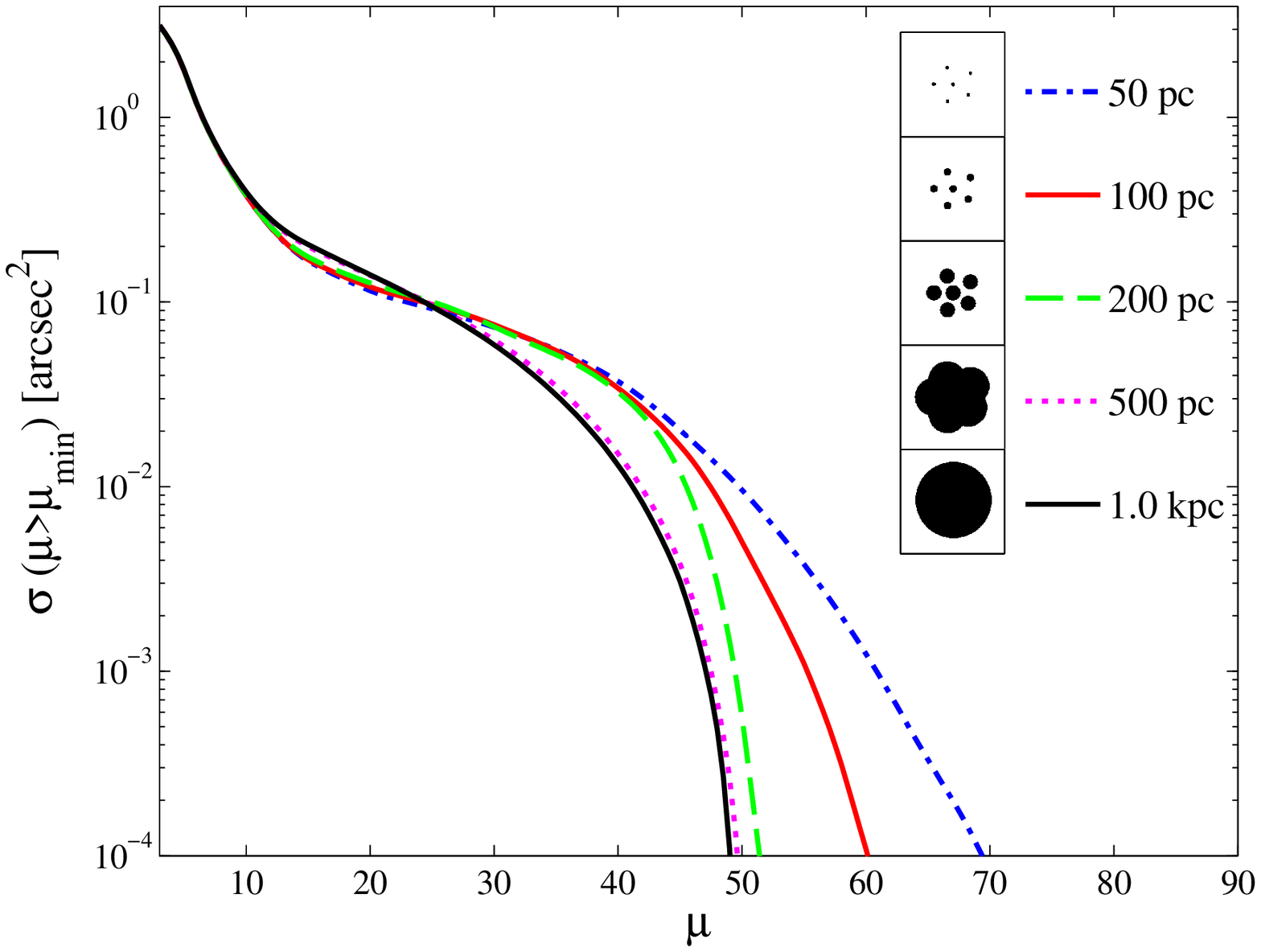}}
\end{minipage}
\caption{ Cumulative magnification cross-sections of different source morphologies (insets) for a $7\times 10^{11} M_{\odot}$ galaxy (mass within the Einstein radius) at $z=0.5$. The source is placed at $z_s=2.0$. The sizes in the legend correspond to the radius of the star forming clumps. All the insets are on the same scale.
\textbf{Left:} Source models with varying numbers of clumps, to fill the same sky area as the 1~kpc source. 
\textbf{Right:} Source models with a fixed number of clumps in fixed locations. 
}

\label{f:rclumpiness}
\end{figure*}

\section{Results}
\label{sec:results}

We applied the models to investigate two related but separate effects: 1) given some
dispersion in intrinsic source sizes, how does selecting the brightest lenses
affect the inferred source size distribution; 2) for a single multi-component source
(modeled here as one small emission region embedded in a larger emission region), how does lensing selection affect the
relative amount of flux received from regions of different sizes?

\subsection{Size Selection}
\label{sec:size}

The modeling shown in Figure~\ref{f:sizes} demonstrates that the distribution of source sizes derived from samples of lensed 
objects are unlikely to be representative of the underlying source size distribution.
Starting from a population of sources that are equally distributed among four sizes (Section~\ref{sec:popmodels}), we find that 
the selection of millimeter-bright sources ($>10$~mJy at 1.4~mm) introduces a strong selection bias toward the 
most compact sources. 
Compact sources can therefore be expected to be overrepresented by a factor of two or more in a survey with a high flux 
density threshold, while larger sources would be nearly completely missed.

This result is not 
trivial, as it represents the balance between the higher total magnification achievable for a compact source and the lower probability of
aligning the compact source with a lensing caustic to achieve high magnification. 
Figure~\ref{f:sizes} demonstrates that the bias persists across a range of realistic lens ellipticities, with a stronger bias toward compact objects for more elliptical lensing potentials. The ellipticities of real lenses will be distributed through and beyond the $\epsilon=0.1-0.3$ range shown, and this unknown distribution function would be required in order to predict a
robust correction for the size bias in observed populations.

\begin{figure}[h]
\centering
\makebox[0cm]{\includegraphics[width=8.5cm]{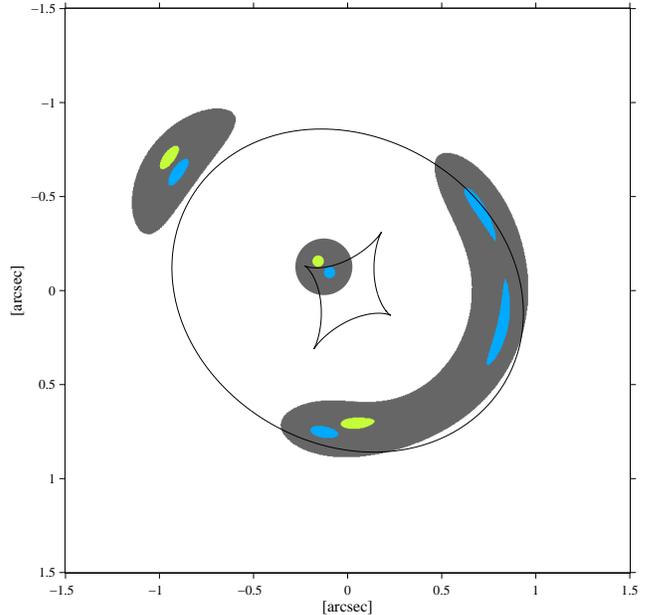}}
\caption{A lens model showing the effects of differential magnification. The total magnification of the three subcomponents are 11.7, 17.1, and 5.2 for the extended source (gray) compact source inside the caustic (blue), and the doubly imaged source (light green) respectively. Here the second compact component is assumed in order to emphasize the strong dependence of magnification on the exact spatial configuration of the components. }
\label{f:image}
\end{figure}

The preference for compact sources at the highest flux densities can be explained by examining the distribution of 
magnification in the lensed population as a function of observed flux density (Figure~\ref{f:dpdmu}). At larger observed flux densities, 
the distribution of source magnifications in the lensed sample (unlensed sources are excluded) shifts to larger values. 
The distribution of magnifications is dependent on many aspects of the source population model and the distribution of lensing 
halo shapes and masses. However, the basic conclusion that sources of the highest observed flux density are dominated by 
high-magnification sources is robust. The brightest sources are drawn from a wide range of intrinsic fluxes or, equivalently, objects 
of the same observed flux will derive from a wide range in $\mu$.
A sample of sources with a high flux density threshold will best identify very high magnification sources, which also provides the highest source-plane spatial resolution for studying the ISM on small scales in these distant galaxies.

We also study the possible effect of the clumpiness of the source light profile on lensed number counts. High-spatial resolution observations 
have shown that luminous SMGs are composed of a few star forming knots \citep{Swinbank:10b}. 
Since each knot is smaller in size than the overall extent of the galaxy, it is possible that the probability of achieving high magnification 
will change significantly with the size of the knot.
To study this effect we calculate the magnification cross-section, the solid angle in the lens plane over which the magnification 
will exceed a given value, for a source of roughly fixed size (1~kpc radius) but broken into clumps of five different sizes, radii of 1~kpc, 
and 500, 200, 100, and 50 pc. 

In Figure~\ref{f:rclumpiness}, we compare the lensing cross-section for the various source morphologies.
In the left panel, where for each clump size we distribute them over the entire region covered by the 1~kpc source, 
we find that the clumpiness of the emission has very little impact on the overall lensing magnification. 
The clumps sample the same region of the lens as the single source, so the magnification averaged over the 
ensemble of clumps is similar to that for the single source. 
In the right panel, we divide the 1~kpc source into just six clumps of different sizes. The sparser sampling of the 
lens leads to small but measurable disparities in the magnification cross section between the models with different clump sizes. 
The smallest clumps have 
the largest possible magnifications, but also have a reduced cross section between $\mu=10$ and 20 for this combination 
of clump and lens geometry. Together, these tests demonstrate that the total extent of the background galaxy, rather than its 
division into clumps, is primarily responsible for determining how likely it is to be lensed, with the small scale structure having some impact 
on the total magnification.

\begin{figure*}[h]
\begin{center}
\centering
\begin{minipage}[t]{0.45\linewidth}
\centering
\includegraphics[width=1.00\textwidth]{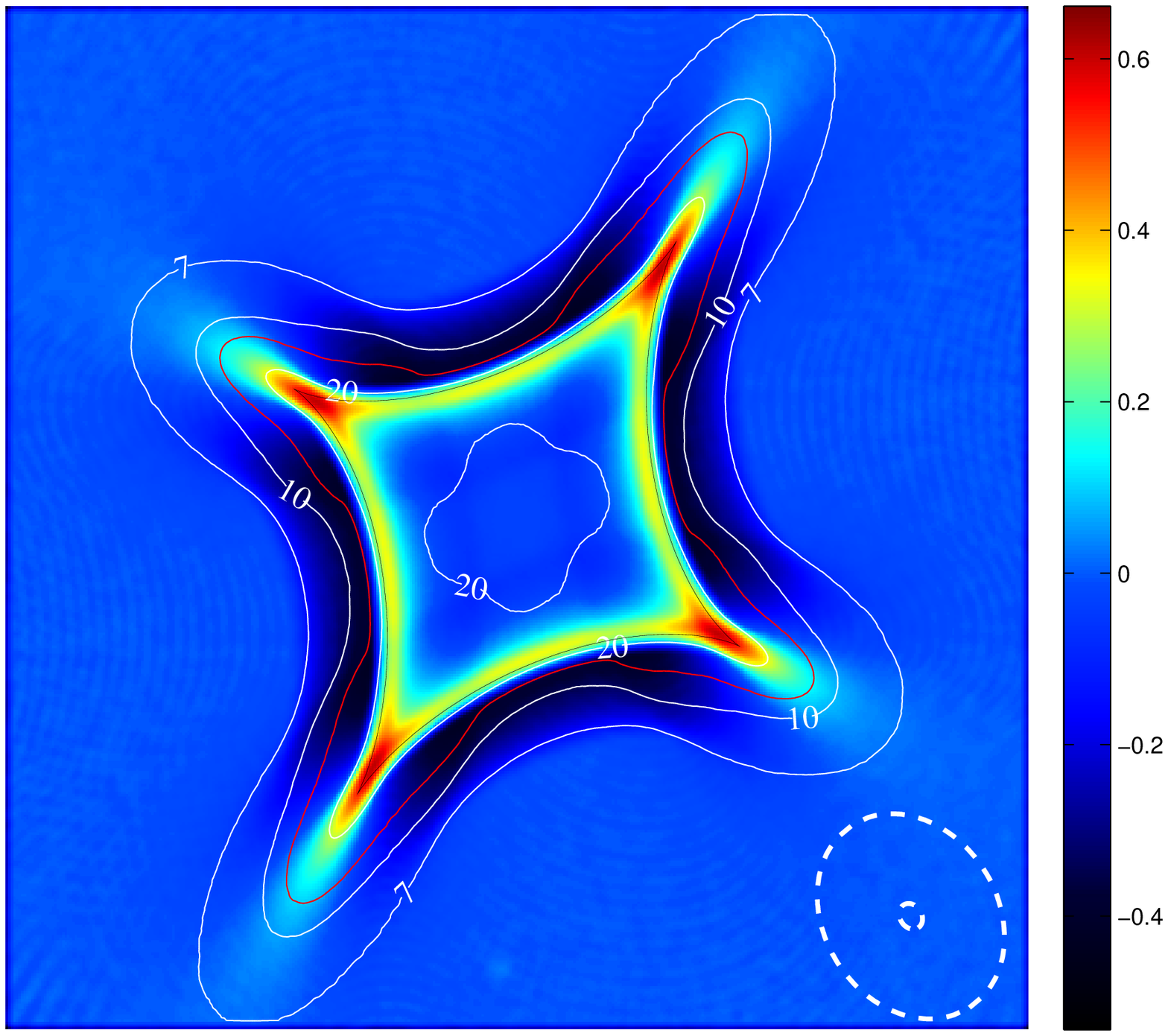}\\
\end{minipage}
\begin{minipage}[t]{0.45\linewidth}
\centering
\includegraphics[width=1.00\textwidth]{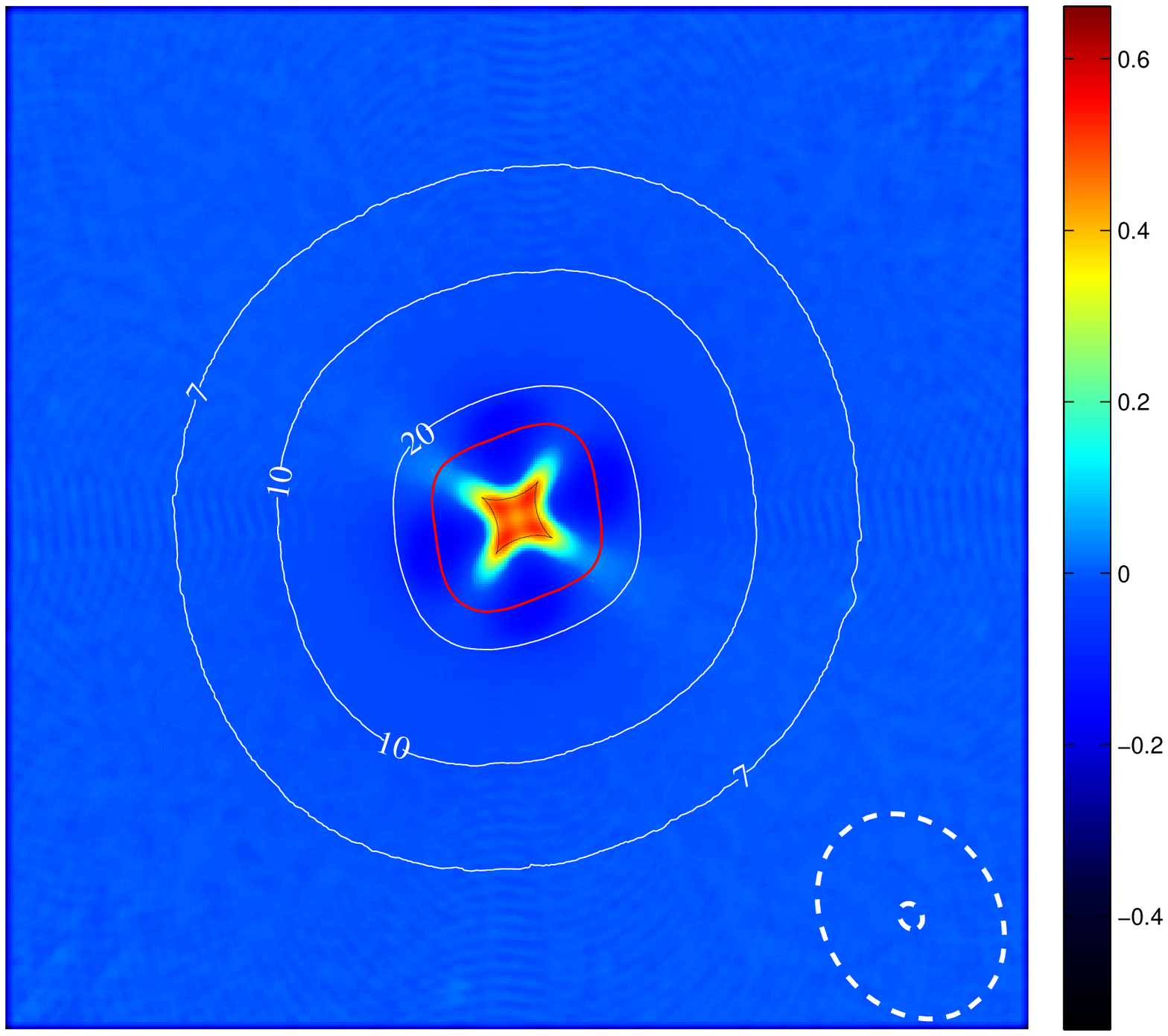}\\
\end{minipage}

\begin{minipage}[t]{0.45\linewidth}
\centering
\includegraphics[width=1.00\textwidth]{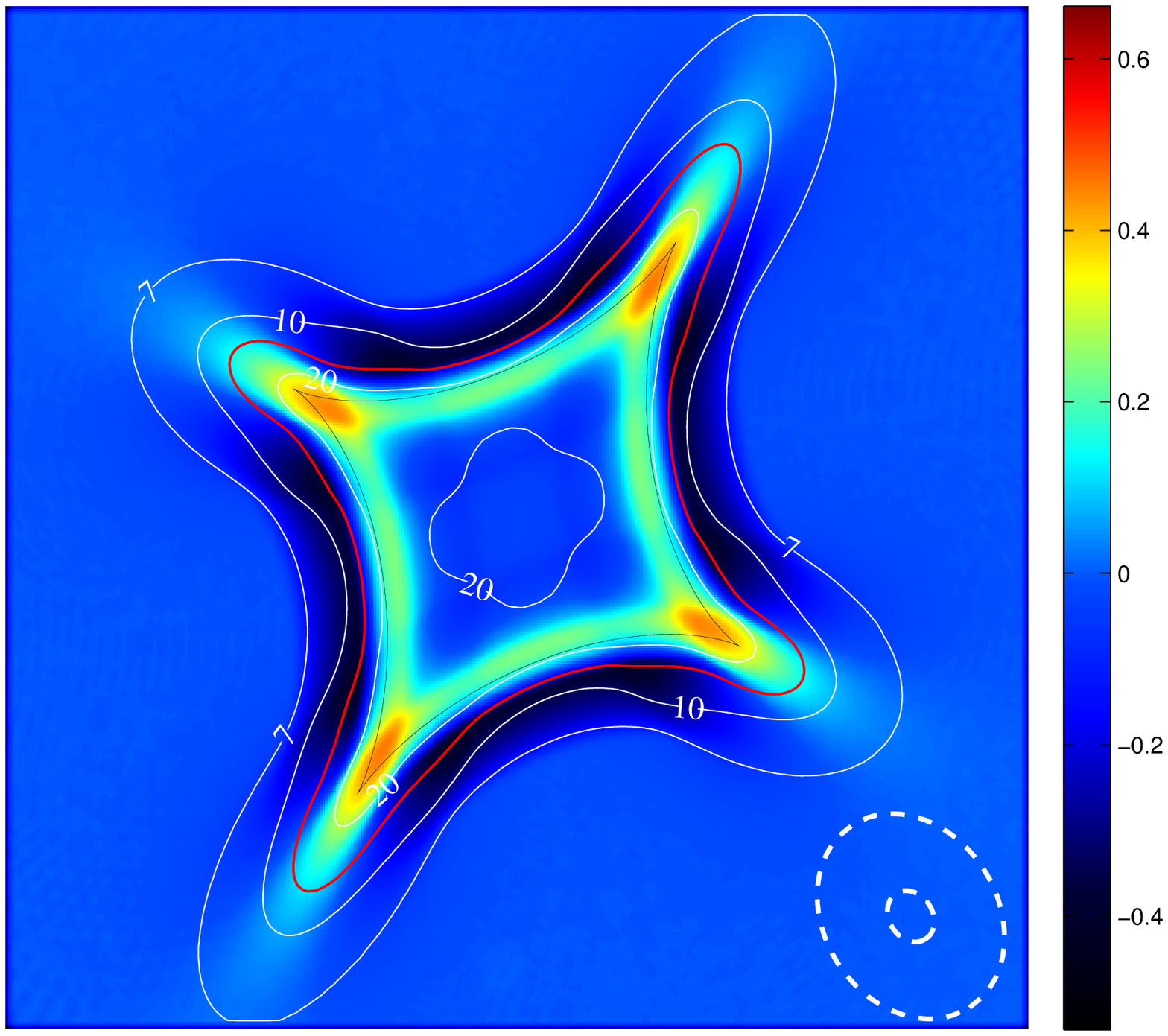}\\
\end{minipage}
\begin{minipage}[t]{0.45\linewidth}
\centering
\includegraphics[width=1.00\textwidth]{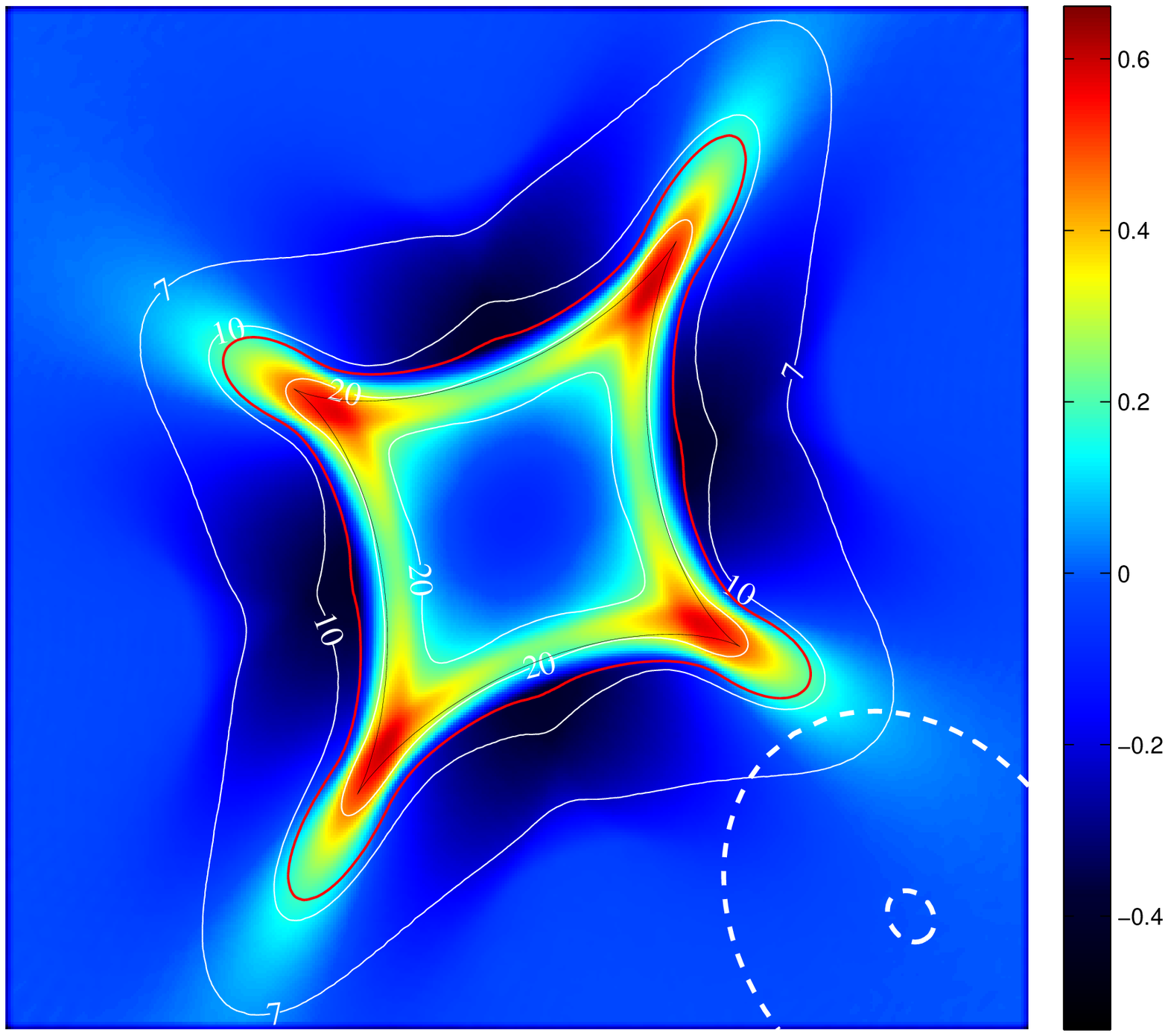}\\
\end{minipage}
\end{center}
\caption{\label{f:variations} 
{\it Top left}: Magnification ratio as a function of position in the source plane, with $\epsilon=0.3$, $\alpha_\mathrm{R}=R_\mathrm{core}/R_\mathrm{ext} = 1/8$, and intrinsic flux ratio $\beta_\mathrm{F}=1$; this is our fiducial model.
The total magnification is indicated in all panels with white contours. The red countour traces the line of constant total magnification that passes through the minima in the magnification ratio ($\beta_{\rm F}'$).
The source model is outlined in the lower right.
{\it Top right}: The fiducial model, but with lens ellipticity decreased ($\epsilon=0.1$).
{\it Bottom left}: The fiducial model, but with the source geometry altered ($\alpha_\mathrm{R} = 1/4$).
{\it Bottom right}: The fiducial model, but with the source size doubled.}
\end{figure*}

\subsection{Differential Magnification}
The selection of high-magnification sources assures close alignment between the lensing caustics and regions of bright emission on the SMG.
This increases the importance of the differential magnification effect, which is demonstrated in Figure~\ref{f:image}. Here, adjacent compact regions 
that straddle the tangential caustic differ in magnification by a factor of three, while the total magnification of the extended component that 
surrounds them is comparable to  the magnification of the brightest component. For real sources that are a composite of multiple components of 
significantly varying conditions, the effects of differential magnification on the observed SED and spectral line ratios can be dramatic \citep{serjeant:12}.

\begin{figure}[h]
\centering
\makebox[0cm]{\includegraphics[width=8.5cm]{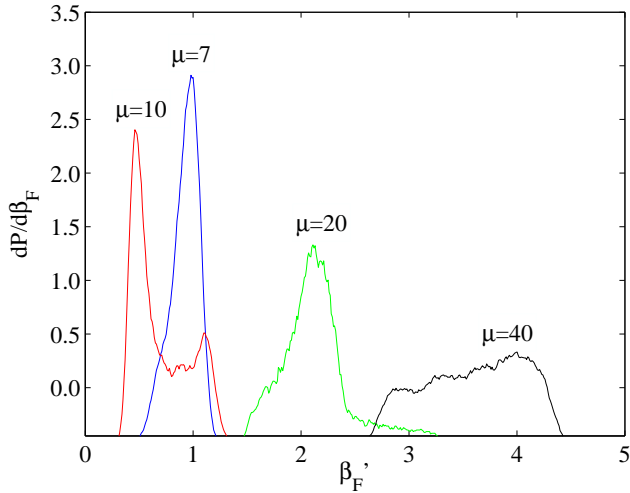}}
\caption{Normalized distribution of flux ratios at each magnification. At intermediate magnifications (e.g. $\mu\sim10$) there is a higher probability of a larger magnification for more extended objects. At higher magnifications the situation is reversed. This combined with the overall magnification distribution at an observed flux (Figure \ref{f:dpdmu}) results in a serious difficulty in interpreting the intrinsic flux ratio of different components.}
\label{f:ratios}
\end{figure}

Starting from the simple source structure described in Section~\ref{sec:smgmodel}, we consider lensing effects for several galaxy and lens
variations in Figure~\ref{f:variations}. A fiducial model with $\epsilon=0.3$, $\alpha_\mathrm{R}= 1/8$, and intrinsic flux ratio 
$\beta_\mathrm{F}=1$ is shown in the top left. The source plane is shown, the color scale represents the ($log_{10}$) observed flux ratio, corresponding to the ratio
of magnifications ($\beta_\mathrm{F}'$) of the two components as a function of the source-lens alignment. The ratio varies by a factor $>10$
across this image, including substantial area over which the extended component can be accentuated by a factor of several.

The lens and source parameters both strongly affect the observed ratios. The top left and right panels show that in a less elliptical lens,
the magnification ratio varies in a smaller region. The similarity between the size of the diamond caustic and the ``core'' emission component
results in less variation in the magnification ratio across the source. The left panels demonstrate that making the sizes of the two components 
more similar decreases the variations in magnification ratio. Increasing the size of the whole source (lower right) slows the rate of change of the
magnification ratio. The red contour in each plot shows the curve of constant $\mu_\mathrm{T}$ that connects the points of minimum 
$\beta_\mathrm{F}'$. This curve is always very close to the caustic, where slight changes in source-lens alignment can mean dramatic changes 
in magnification ratio.

The probability distribution for $\beta_\mathrm{F}'$ at fixed $\mu_\mathrm{T}$ is shown in Figure~\ref{f:ratios}, for the model in the
top left panel of Figure~\ref{f:variations}. At moderate magnifications, the extended component is often magnified by a larger factor, while 
the highest magnifications over emphasize the core but with a large range in amplitude. Considering Figures~\ref{f:dpdmu} and~\ref{f:variations} together, we see that for a sample selected with a high flux density limit, the distortion of the SED cannot be
assumed to be understood.

A more optimistic view of the situation can be taken from Figure~\ref{f:variations}. A good lens model can greatly reduce the uncertainty
in the magnification distribution across the source. Regions where the flux is dominated by the diffuse component are
confined to those areas just outside the ``diamond caustic'' that are well-separated
from the cusps. High-resolution imaging with ALMA or \textit{Hubble~Space Telescope} will provide lens models that will
be useful for these purposes. While it would be preferable to have high-resolution imaging
at all frequencies, for simple questions about whether the flux is dominated by the
compact or diffuse regions it could be sufficient to have a good lens model at
a single frequency. More advanced inferences are available with multiple wavelengths of resolved imaging
or by assuming a physically motivated size structure \citep[e.g.,][]{Blain:99}.

\section{conclusion}
\label{sec:concl}
Strongly lensed SMGs permit close examination of these rapidly star-forming galaxies with short integration times. 
However, the selection of lensed objects and the strong lensing itself introduce biases that must be considered when 
interpreting the observed galaxy properties.

We have demonstrated that strongly lensed SMGs selected by a flux cut in wide surveys are affected by a ``size bias''. They are more likely to include more compact galaxies, which correspond to higher surface brightness sources at a given unlensed flux. More specifically, as one probes samples with higher observed flux the distribution of the size of the sources is increasingly skewed toward more compact objects.

We have also studied the effects of differential lensing of SMGs. We model the SMGs as having two components, a compact core embedded in a more extended region, and measure the relative flux of these components in the presence of strong lensing.
 We find that the relative gain for compact and extended components depends sensitively on the source-lens alignment. At high total magnifications the observed emission from the compact core is amplified by a larger factor relative to the emission from the diffuse region. At intermediate magnifications (e.g., $\mu \sim 10$), the situation is often reversed with the diffuse emission being magnified by a larger factor.
 A similar range of effects can be expected for spectral lines. 

This suggests that without applying the required corrections from lens models, the SED and the ratios of molecular lines could be significantly biased. Fortunately, the relative magnification of the compact and extended regions depends on the geometry of the caustics and the source-lens alignment in a highly regular manner. This effect can therefore be well understood by constructing reliable lens models for each source, which will require spatially resolved imaging data.

\acknowledgements{This work was
supported by FQRNT International Internship grant, as well as the Canadian Institute for Advanced Research, NSERC Discovery and the Canada Research Chairs program. G.H. acknowledges the
hospitality of KICP (Chicago) and FNAL. 
Y.D.H. acknowledges the hospitality of IoA (Cambridge) and Caltech during his visits and thanks Chris Carrili, Andrew Blain, and Joaquin Vieira for useful discussions. As this work neared completion we became aware of another manuscript \citep{serjeant:12} that investigates related issues.}

\bibliographystyle{apj}
\bibliography{references}

\end{document}